\documentclass[prl,onecolumn,aps,showpacs,floats,letterpaper,floatfix,groupedaddress,preprintnumbers,nofootinbib]{revtex4}

\usepackage{graphicx,epsf, epsfig, amssymb}% Include figure files
\usepackage{bm}
\usepackage{longtable}

\def\be{\begin{equation}}
\def\ee{\end{equation}}
\def\beq{\begin{eqnarray}}
\def\eeq{\end{eqnarray}}

\begin{document}

\title{Stability of naked singularities and algebraically special modes}

\author{Vitor Cardoso} \email{vcardoso@phy.olemiss.edu}
\affiliation{Department of Physics and Astronomy, The University of
Mississippi, University, MS 38677-1848, USA \footnote{Also at Centro
de F\'{\i}sica Computacional, Universidade de Coimbra, P-3004-516
Coimbra, Portugal}}

\author{Marco Cavagli\`a} \email{cavaglia@phy.olemiss.edu}
\affiliation{Department of Physics and Astronomy, The University of Mississippi, University, MS
38677-1848, USA}

\date{\today}

\begin{abstract}
We show that algebraically special modes lead to the instability of naked
singularity spacetimes with negative mass. Four-dimensional negative-mass
Schwarzschild and Schwarzschild-de Sitter spacetimes are unstable. Stability of
the Schwarzschild-anti-de Sitter spacetime depends on boundary conditions. We
briefly discuss the generalization of these results to charged and rotating
singularities.
\end{abstract}
%%%%%%%%%%%%%%%%%%%%%%%%%%%%%%%%%%%%%%%%%%%%%%%%%%%%%%%%%%%%%%%%%%%%%%%%%%%%%%%%%%%%%%%%%%%%%%%%%%%%%%%%%%%%%%%
\maketitle
%%%%%%%%%%%%%%%%%%%%%%%%%%%%%%%%%%%%%%%%%%%%%%%%%%%%%%%%%%%%%%%%%%%%%%%%%%%%%%%%%%%%%%%%%%%%%%%%%%%%%%%%%%%%%%%
%%%%%%%%%%%%%%%%%%%%%%%%%%%%%%%%%%%%%%%%%%%%%%%%%%%%%%%%%%%%%%%%%%%%%%%%%%%%%%%%%%%%%%%%%%%%%%%%%%%%%%%%%%%%%%%%%%%
\section{Introduction}
%%%%%%%%%%%%%%%%%%%%%%%%%%%%%%%%%%%%%%%%%%%%%%%%%%%%%%%%%%%%%%%%%%%%%%%%%%%%%%%%%%%%%%%%%%%%%%%%%%%%%%%%%%%%%%%%%%%
Discussion of naked singularities has recently attracted some attention in the literature
\cite{Gibbons:2004au,Gleiser:2006yz}. The cosmic censorship conjecture of general relativity states that
spacetimes with naked singularities do not exist in nature because the evolution of regular initial data set
must either form smooth geometries or give rise to singularities surrounded by horizons. However, the cosmic
censorship conjecture could be far a too restrictive requirement for the non-existence of these spacetimes.
Even if they form, naked singularity solutions can still be unphysical provided they are unstable under
perturbations. Stability of spacetimes with naked singularities is thus a subject worth investigating as an
alternative (or complement) to the cosmic censorship conjecture. In this paper we focus on naked singularities
arising from negative-mass solutions of four-dimensional Einstein gravity coupled to a cosmological constant,
namely the Schwarzschild-(Anti)-de Sitter (S(A)dS) geometries. Although these spacetimes cannot be obtained
through gravitational collapse of matter sources with positive energy, their perturbative analysis is
relatively simple and provides a treatable toy model for the investigation of the stability/instability regimes
of naked singularities.

Previous work on the subject includes that of Gibbons, Hartnoll and Ishibashi \cite{Gibbons:2004au}, who first
considered the stability of negative-mass spacetimes. The stability of these geometries can be reduced to the
study of two classes of perturbations, the odd or Regger-Wheeler perturbations and the even or Zerilli
perturbations. It was shown that negative-mass Schwarzschild spacetimes are stable against both kinds of
perturbations. However, as was pointed out by Gleiser and Dotti \cite{Gleiser:2006yz}, the analysis of the even
perturbations of Ref.\ \cite{Gibbons:2004au} is flawed. They argue that the marginally stable solution used to
prove stability is at most twice differentiable in the radial coordinate. Thus the perturbed metric is not
continuous and is not a solution of the linearized vacuum Einstein equations. (It actually solves the field
equations with a singular distribution of sources.) Moreover, boundary conditions at the singularity cannot be
chosen by self-adjoint extensions of the wave operator for gravitational perturbations \cite{wald} (see Ref.\
\cite{Gleiser:2006yz} and discussion below). The choice of boundary conditions of Ref.\ \cite{Gibbons:2004au}
is not consistent with the perturbation expansion because it requires the radial components of the even
perturbations to diverge near the singularity. These two points led Gleiser and Dotti to conclude that the
negative-mass Schwarzschild spacetime is unstable against even perturbations and derive an explicit solution
for the unstable mode. We will show that this solution is an algebraically special mode \cite{chandra} of the
geometry and extend this result to the S(A)dS spacetimes. The method presented below also provides a framework
to address the stability of more general negative-mass spacetimes. For example, our conclusions are easily
generalized to charged Reissner-Nordstr\"{o}m and rotating Kerr geometries.

%%%%%%%%%%%%%%%%%%%%%%%%%%%%%%%%%%%%%%%%%%%%%%%%%%%%%%%%%%%%%%%%%%%%%%%%%%%%%%%%%%%%%%%%%%%%%%%%%%%%%%%%%%%%%%%%%%%
\section{Spacetime perturbations and algebraically special modes}
%%%%%%%%%%%%%%%%%%%%%%%%%%%%%%%%%%%%%%%%%%%%%%%%%%%%%%%%%%%%%%%%%%%%%%%%%%%%%%%%%%%%%%%%%%%%%%%%%%%%%%%%%%%%%%%%%%%
Gravitational perturbations in spherically symmetric, static geometries \cite{regge,Zerilli:1970se} are divided
in two distinct classes according to their transformation properties: Regge-Wheeler or odd perturbations ($+$)
and Zerilli or even pertubations ($-$). These pertubations are defined by assuming that the metric functions
can locally be written as
\begin{equation}
g_{\mu \nu}({\bf x})= g^{(0)}_{\mu \nu}({\bf x})+h_{\mu\nu}({\bf x})\,, \label{perturb}
\end{equation}
where $\mu$ and $\nu$ take values between $0$ and $3$ and $\bf{x}$ is a vector of spacetime coordinates. The
metric $g^{(0)}_{\mu \nu}(\bf{x})$ is the static background solution of Einstein's equations and $h_{\mu
\nu}(\bf{x})$ is a small perturbation. In the case under consideration, the background metric is
\begin{equation}
ds^{2}= fdt^{2}- f^{-1}dr^{2}-r^{2}(d\theta^{2}+\sin^2\theta d\phi^{2})\,,\qquad
f(r)=\left(\alpha\frac{r^{2}}{R^2}+1-\frac{2M}{r}\right)\,,
\label{lineelement}
\end{equation}
where $\alpha=+1$ for AdS, $-1$ for dS and $R$ is the (A)dS radius. The quantity $M$ is the mass of the
spacetime, which we allow to be positive or negative. Substituting Eq.\ (\ref{perturb}) in the field equations,
we obtain the differential equations for the perturbations. We use the original notation of Regge and Wheeler
\cite{regge}. (See Refs.\ \cite{Cardoso:2001bb,Cardoso:2003cj} for details.) The perturbations are expanded in
tensorial spherical harmonics with parity $(-1)^{l+1}$ (odd) and $(-1)^l$ (even), where $l$ is the angular
momentum of the mode. Choosing the Regge-Wheeler gauge, the canonical form for the perturbations is
\begin{itemize}
\item odd parity:
\begin{eqnarray}
h_{\mu \nu}= \left[
 \begin{array}{cccc}
 0 & 0 &0 & h_0(r)
\\ 0 & 0 &0 & h_1(r)
\\ 0 & 0 &0 & 0
\\ h_0(r) & h_1(r) &0 &h_0(r)
\end{array}\right] e^{-i \omega t}
\left(\sin\theta\frac{\partial}{\partial\theta}\right) P_l(\cos\theta)\,,
\label{odd}
\end{eqnarray}
\item even parity:
\begin{eqnarray}
h_{\mu \nu}= \left[
 \begin{array}{cccc}
 H_0(r) f(r) & H_1(r) &0 & 0
\\ H_1(r) & H_2(r)/f(r)  &0 & 0
\\ 0 & 0 &r^2K(r) & 0
\\ 0 & 0 &0 & r^2K(r)\sin^2\theta
\end{array}\right] e^{-i \omega t}
P_l(\cos\theta)\,,
\label{even}
\end{eqnarray}
\end{itemize}
where $P_l(\cos\theta)$ is the Legendre polynomial with angular momentum $l\geq 2$. (We consider only radiative
multipoles). The functions $h_0(r)\,, h_1(r)\,, H_0(r)\,,H_1(r)$ and $K(r)$ are functions of the radial
coordinate $r$ alone. For more details on this decomposition we refer the reader to the original works
\cite{regge,Zerilli:1970se}. Inserting Eq.\ (\ref{odd}) and Eq.\ (\ref{even}) into Einstein's equations, one
obtains ten coupled second-order differential equations for the perturbations. These equations can be combined
into two second-order differential equations, one for each parity \cite{regge, Zerilli:1970se}. The
perturbations can be written in terms of a unique wave function $\Psi(r)$. For odd parity, we have
\begin{equation}
\Psi_+(r)=  \frac{f(r)}{r} h_1(r)\,,\qquad
h_0=\frac{i}{\omega}\frac{d}{dr_*}\left(r\Psi_+ \right)\,.
\label{qodd}
\end{equation}
Einstein's equations reduce to
\begin{equation}
\frac{\partial^{2} \Psi_+}{\partial r_*^{2}} + \left\lbrack\omega^2 -V_{+}(r)\right\rbrack \Psi_+=0 \,,
\label{waveodd}
\end{equation}
where
\begin{equation}
V_{+}=  f(r) \left\lbrack\frac{l(l+1)}{r^2}-\frac{6M}{r^3}\right\rbrack
\label{vodd}
\end{equation}
and the tortoise coordinate $r_*$ is defined by $dr/dr_*=f(r)$. For even parity,
$H_0$, $H_1$ and $K$ are related to $\Psi_-(r)$ by
\begin{eqnarray}
K&=& \frac{6M^2+c\left(1+c\right)r^2+M\left(3cr-3\alpha\frac{r^3}{R^2}\right)} {r^2\left(3M+cr\right)} \Psi_-
+\frac{d\Psi_-}{dr_*} \,, \\
H_1&=& -\frac{i\omega\left(-3M^2-3cMr+cr^2-3M\alpha\frac{r^3}{R^2}\right)} {r \left(3M+cr\right) f(r)} \Psi_-
-i \omega \frac{r}{f(r)}\frac{d\Psi_-}{dr_*}\,, \label{4.10}
\end{eqnarray}
where $c=\frac{1}{2}\left\lbrack l(l+1)-2\right\rbrack$. Einstein's equations
are
\begin{equation}
\frac{\partial^{2} \Psi_-}{\partial r_*^{2}} + \left\lbrack \omega^2 -V_{-}(r)\right\rbrack \Psi_-=0 \,,
\label{waveeven}
\end{equation}
where
\begin{equation}
V_{-}= \frac{2f(r)}{r^3}
\frac{9M^3+3c^2Mr^2+c^2\left(1+c\right)r^3+3M^2\left(3cr+3\alpha\frac{r^3}{R^2}\right)} {\left(3M+cr\right)^2}
\,. \label{veven}
\end{equation}
The potentials $V_\pm$ can be rewritten in the form \cite{chandra}
\begin{equation}
V_\pm=W^2\pm\frac{dW}{dr_*}+\beta\,,
\label{V2}
\end{equation}
where \footnote{The following equations correct some typos in Ref.\ \cite{Cardoso:2001bb}.}
\begin{eqnarray}
W&=&\frac{3M\left (6Mr^2+2c\alpha R^2(2M-r) \right )}{cr^2\left (6M+2cr\right )\alpha R^2}+j\,,\\
\beta&=&-\left (\frac{(l-1)l(l+1)(l+2)}{12M}\right )^2\,,\\
j&=&-\frac{(l-1)l(l+1)(l+2)}{12M}-\frac{3M\alpha}{cR^2}\,.
\end{eqnarray}
Potentials of the form (\ref{V2}) are sometimes called superpartner potentials
\cite{Cooper:1994eh}. A class of special modes can be found analytically using
Eq.\ (\ref{V2}) \cite{chandra}. Defining
\begin{equation}
\pm W=\frac{d}{dr_*}\log \chi_\pm\,,
\end{equation}
the wave equations can be written as
\begin{equation}
\frac{1}{\Psi_{\pm}}\frac{d^2\Psi_{\pm}}{dr_*^2}+\left (\omega^2-\beta \right )=
\frac{1}{\chi_\pm}\frac{d^2\chi_\pm}{dr_*^2}\,, \label{equationgeneral}
\end{equation}
where we have used the identity
\begin{equation}
\frac{1}{\chi_\pm}\frac{d^2\chi_\pm}{dr_*^2}=W^2\pm\frac{dW}{dr_*}\,.
\end{equation}
The algebraically special modes are defined by $\omega^2-\beta=0$. In this
case, Eq.\ (\ref{equationgeneral}) can be integrated. Its general solution is
\begin{equation}
\chi_\pm=\exp\left[\pm\int W dr_*\right]\,,\qquad \Psi_{\pm}=\chi_\pm \int \frac{dr_*}{\chi_\pm^2}=\chi_\pm\left
(C_1+C_2\int_{0}^r \frac{dr_*}{\chi_\pm^2}\right )\,. \label{sol}
\end{equation}
%
%%%%%%%%%%%%%%%%%%%%%%%%%%%%%%%%%%%%%%%%%%%%%%%%%%%%%%%%%%%%%%%%%%%%%%%%%%%%%%%%%%%%%%%%%%%%%%%%%%%%%%%%%%%%%%%%%%%
\section{Instability of negative-mass spacetimes}
%%%%%%%%%%%%%%%%%%%%%%%%%%%%%%%%%%%%%%%%%%%%%%%%%%%%%%%%%%%%%%%%%%%%%%%%%%%%%%%%%%%%%%%%%%%%%%%%%%%%%%%%%%%%%%%%%%%
Algebraically special modes can be used to prove the instability of the S(A)dS, spherically symmetric
negative-mass spacetimes. Since the singularity at $r=0$ is naked, the wave equations have support in the
interval $r\in(0,r_c)$, where $r_c$ is infinity for Schwarzschild and SAdS spacetimes, and the cosmological
horizon for SdS spacetime. It can be shown that the algebraically special modes with frequency
\begin{equation}
\omega_{\rm s}= i \frac{(l-1)l(l+1)(l+2)}{12|M|}
\end{equation}
grow exponentially with time, i.e.\ they are unstable despite being regular
throughout the spacetime. The behavior of negative-mass perturbations for $r\to
0$ is
\begin{equation}
\Psi_+ \sim \frac{C_{+1}}{r}+C_{+2}r^3\,,\qquad
\Psi_- \sim C_{-1}r+C_{-2}r\log{r}\,.
\label{beh0}
\end{equation}
The logarithm term makes the perturbation quantity $K$ diverge at the origin.
Since we are looking for spatially bounded metric perturbations, we require
\begin{equation}
C_{+1}=C_{-2}=0\,.
\label{constants}
\end{equation}
In addition to the boundaries $r=0$ and $r=r_c$, the potentials $V_\pm$ seem to
diverge at
\begin{equation}
r=r_i\equiv\frac{6|M|}{(l-1)(l+2)}\,.
\end{equation}
Near this point the wave function behaviors are
\begin{equation}
\Psi_+ \sim C_{+2}\times{\rm constant}\,,\qquad
\Psi_- \sim \frac{C_{-1}}{r-r_i}\,,
\end{equation}
where we have used Eq.\ (\ref{constants}). The odd-parity wave function
$\Psi_+$ is well-behaved for $r\to r_i$. The even-parity wave function $\Psi_-$
diverges. However, from Eq.\ (\ref{4.10}) it follows
\begin{equation}
K\sim\left (1+\frac{2c}{3}+\frac{9M^2\alpha}{c^2R^2}\right )\left ( \frac{\Psi_-}{r-r_i} +\frac{d\Psi_-}{dr}
\right )\,, \qquad H_1\sim -i\omega r_i\left ( \frac{\Psi_-}{r-r_i} +\frac{d\Psi_-}{dr} \right )\,.
\label{pertnearri}
\end{equation}
Therefore, both perturbations are regular at $r=r_i$. To complete the discussion
and assess the stability of the spacetime, we still have to consider spatial
infinity.
%%%%%%%%%%%%%%%%%%%%%%%%%%%%%%%%%%%%%%%%%%%%%%%%%%%%%%%%%%%%%%%%%%%%%%%%%%%%%%%%%%%%%%%%%%%%%%%%%%%%%%%%%%%%%%%%%%%
\subsection{Schwarzschild spacetime}
%%%%%%%%%%%%%%%%%%%%%%%%%%%%%%%%%%%%%%%%%%%%%%%%%%%%%%%%%%%%%%%%%%%%%%%%%%%%%%%%%%%%%%%%%%%%%%%%%%%%%%%%%%%%%%%%%%%
In the asymptotically flat case, $r_* \sim r$ for large $r$. Equation
(\ref{sol}) and Eq.\ (\ref{constants}) imply
\begin{equation}
\chi_\pm \sim e^{\pm |\omega_{\rm s}|r}\,, \qquad
\Psi_+=C_{+2} \chi_+ \int_0^{r}\frac{1}{\chi_+^2}dr_*\,,\qquad
\Psi_-=C_{-1}\chi_-\,.
\label{psi-flat}
\end{equation}
The integral above converges and is positive-definite. The behavior of the
odd-parity wave function for large $r$ is
\begin{equation}
\Psi_{+}\sim C_{+2}~ e^{|\omega_{\rm s}|r}\,.
\end{equation}
The wave function $\Psi_+$ is bounded only if $C_{+2}=0$. Odd-parity perturbations
are thus stable at this algebraically special frequency. This conclusion agrees
with the results of Ref.\ \cite{Gibbons:2004au}, where Regge-Wheeler modes are
shown to be stable for all frequencies. Equation (\ref{psi-flat}) also shows that
$\Psi_-$ is bounded on the entire space. However, the even-parity mode grows
exponentially in time as $e^{-i\omega t}= e^{{\rm Im}\left(\omega_{\rm s}\right)
t}$. We conclude that the negative-mass Schwarzschild spacetime is unstable against
the algebraically special Zerilli modes. The explicit form of the perturbations is
given by
\begin{equation}
\Psi_{-}=C_{-1}\left(\frac{r}{6M+(l-1)(l+2)r}\right)e^{-|\omega_{\rm s}|r_*}\,.
\end{equation}
The above result agrees with the unstable modes presented by Gleiser and Dotti
\cite{Gleiser:2006yz}.
%%%%%%%%%%%%%%%%%%%%%%%%%%%%%%%%%%%%%%%%%%%%%%%%%%%%%%%%%%%%%%%%%%%%%%%%%%%%%%%%%%%%%%%%%%%%%%%%%%%%%%%%%%%%%%%%%%%
\subsection{SdS spacetime}
%%%%%%%%%%%%%%%%%%%%%%%%%%%%%%%%%%%%%%%%%%%%%%%%%%%%%%%%%%%%%%%%%%%%%%%%%%%%%%%%%%%%%%%%%%%%%%%%%%%%%%%%%%%%%%%%%%%
The behavior of the negative-mass SdS geometry near the cosmological horizon
$r_c$ is
\begin{equation}
f=-\frac{r^2}{R^2}+1+\frac{2|M|}{r}\sim |a|(r_c-r)\,,
\end{equation}
where $a$ is a constant. It follows
\begin{equation}
\chi_\pm \sim e^{\mp |\omega_{\rm s}|\log(r_c-r)}\,.
\end{equation}
Odd-parity modes are stable at the algebraically special frequency. The
algebraically special even-parity modes are regular in all the domain:
\begin{equation}
\Psi_{-}=C_{-1}\left (\frac{r}{6M+(l-1)(l+2)r}\right )e^{-|\omega_{\rm s}|
r_*}\,.
\end{equation}
Zerilli algebraically special perturbations of negative-mass SdS geometry are
spatially regular unstable solutions.
%%%%%%%%%%%%%%%%%%%%%%%%%%%%%%%%%%%%%%%%%%%%%%%%%%%%%%%%%%%%%%%%%%%%%%%%%%%%%%%%%%%%%%%%%%%%%%%%%%%%%%%%%%%%%%%%%%%
\subsection{SAdS spacetime}
%%%%%%%%%%%%%%%%%%%%%%%%%%%%%%%%%%%%%%%%%%%%%%%%%%%%%%%%%%%%%%%%%%%%%%%%%%%%%%%%%%%%%%%%%%%%%%%%%%%%%%%%%%%%%%%%%%%
From the asymptotic behavior of the SAdS geometry with negative mass, it
follows
\begin{equation}
\chi_\pm \sim e^{\mp |\omega_{\rm s}|/r}\,.
\end{equation}
Regge-Wheeler and Zerilli perturbations asymptote to a constant value at spatial
infinity. Boundary conditions play a key role in the SAdS spacetime. If the
perturbations vanish at the boundary (reflective boundary conditions
\cite{Avis:1977yn,Breitenlohner:1982bm,Burgess:1984ti}), the geometry is stable
because $C_{+2}=C_{-1}=0$ imply vanishing perturbations on the entire spacetime.
Other boundary conditions, such as the mixed boundary conditions by Norman and Moss
\cite{Moss:2001ga} must be studied case by case. These conclusions are expected to
hold also in asymptotically AdS spacetimes with different topologies
\cite{Lemos:1994xp, Lemos:1995cm}, e.g.\ toroidal or cylindrical black holes. (For
a discussion on algebraically special modes of toroidal black holes in AdS, see
Refs.\ \cite{Cardoso:2001vs,Miranda:2005qx}).
%%%%%%%%%%%%%%%%%%%%%%%%%%%%%%%%%%%%%%%%%%%%%%%%%%%%%%%%%%%%%%%%%%%%%%%%%%%%%%%%%%%%%%%%%%%%%%%%%%%%%%%%%%%%%%%%%%%
\section{Discussion}
%%%%%%%%%%%%%%%%%%%%%%%%%%%%%%%%%%%%%%%%%%%%%%%%%%%%%%%%%%%%%%%%%%%%%%%%%%%%%%%%%%%%%%%%%%%%%%%%%%%%%%%%%%%%%%%%%%%
The above discussion shows that four-dimensional Schwarzschild and S(A)dS spacetimes with naked singularities
are unstable. The odd-parity  and even-parity gravitational perturbations of these geometries are described by
master equations with superpartner potentials. Algebraically special Regge-Wheeler modes are in general stable,
whereas Zerilli modes lead to instabilities. Similarly to positive-mass spacetimes
\cite{MaassenvandenBrink:2000ru}, algebraically special modes seem to describe different physics according to
their parity.

The extension of these results to negative-mass Reissner-Nordstr\"{o}m and Kerr
spacetimes is straightforward. The equations for the Reissner-Nordstr\"{o}m
spacetime can be found in Ref.\ \cite{chandrabook}. Perturbations in the Kerr
geometry are best handled with Teukolsky's formalism \cite{Teukolsky:1973ha}, which
determines the evolution of the Weyl scalars $\Psi_0$ and $\Psi_4$. Algebraically
special modes for this spacetime were defined by Chandrasekhar \cite{chandra}. (See
also Ref.\ \cite{Onozawa:1996ux}). The explicit solution for the algebraically
special mode is
\begin{equation}
\Psi_0=\frac{1}{r^2-2Mr+a^2}P_{+2}S_{+2}\,,\qquad P_{+2}=\left (a_3r^3+a_2r^2+a_1r+a_0 \right )\,,
\label{pertkerr}
\end{equation}
where $a_i$ are constant parameters and the time dependence is $e^{-i\omega t}$. The value $\omega$ at the
algebraically special mode is determined by the vanishing of the Starobinsky constant. It is easy to see that
well-behaved modes exist. For example, Eq.\ (\ref{pertkerr}) is regular at $r=0$. Therefore,  these special
modes determine the instability of negative-mass Kerr-like naked singularities.

The choice of boundary conditions at the singularity is critical, with various authors considering different
boundary conditions. For instance, the boundary conditions of Ref.\ \cite{Gibbons:2004au} rely on self-adjoint
extensions. In that case, the spacetime is found to be stable. We think the issue of boundary conditions is
simpler in this case: the behavior $\Psi_- \sim r\log r$ seems to be forbidden in a well-defined perturbation
problem. The quantity $K(r)$ would behave as $\log{r}/r$, thus diverging at the singularity faster than the
background.

The purpose of this work was to show the connection between algebraically special modes and instability of
negative-mass naked singularities in four dimensions. An interesting topic for further research would be the
generalization of the above results to higher-dimensional spacetimes, where the potentials for gravitational
perturbations seem no longer be related by superpartner relations
\cite{Kodama:2003jz,Ishibashi:2003ap,Kodama:2003kk}.

%%%%%%%%%%%%%%%%%%%%%%%%%%%%%%%%%%%%%%%%%%%%%%%%%%%%%%%%%%%%%%%%%%%%%%%%%%%%%%%%%%%%%%%%%%%%%%%%%%%%%%%%%%%%%%%%%%
\section*{Acknowledgements}
%%%%%%%%%%%%%%%%%%%%%%%%%%%%%%%%%%%%%%%%%%%%%%%%%%%%%%%%%%%%%%%%%%%%%%%%%%%%%%%%%%%%%%%%%%%%%%%%%%%%%%%%%%%%%%%%%%
We would like to thank Sean Hartnoll for useful correspondence. VC acknowledges financial support from FCT
through the PRAXIS XXI program, and from the Funda\c c\~ao Calouste Gulbenkian through the Programa Gulbenkian
de Est\'{\i}mulo \`a Investiga\c c\~ao Cient\'{\i}fica. This work was supported (in part) by the US Department
of Energy under grant DE-FG02-91ER40622.
%%%%%%%%%%%%%%%%%%%%%%%%%%%%%%%%%%%%%%%%%%%%%%%%%%%%%%%%%%%%%%%%%%%%%%%%%%%%%%%%%%%%%%%%%%%%%%%%%%%%%%%%%%%%%%%%%%

\end{document}